\documentclass[12pt]{iopart}
\usepackage{iopams}
\usepackage{graphicx}
\usepackage[usenames]{color}

\begin{document}

\title{Can a microscopic stochastic model explain the emergence of
pain cycles in patients?}

\author{Francesca Di Patti$^{1,2}$ and Duccio Fanelli$^{1,2,3}$}
\address{$ˆ1$ CSDC Centro Interdipartimentale per lo Studio delle Dinamiche
Complesse, via G. Sansone 1, 50019 Sesto Fiorentino, Italy}
\address{$^2$ INFN, sez. Firenze, via G. Sansone 1, 50019 Sesto Fiorentino,
Italy}
\address{$^3$ Dip. Energetica, Florence University, via di Santa Marta 3, 50139
Firenze, Italy}
\ead{f.dipatti@gmail.com}
\ead{duccio.fanelli@unifi.it}
\begin{abstract}
A stochastic model is here introduced to investigate the  molecular
mechanisms which trigger the perception of pain. The action of
analgesic  drug compounds is discussed in a dynamical context, where
the competition with inactive species is explicitly accounted for.
Finite size effects  inevitably perturb the mean-field dynamics:
Oscillations in the amount of bound receptors spontaneously
manifest, driven by the noise which is intrinsic to the system under
scrutiny. These effects are investigated both numerically, via
stochastic simulations and analytically, through a large-size
expansion. The claim that our findings could provide a consistent
interpretative framework to explain the emergence of cyclic
behaviors in response to analgesic treatments, is substantiated.
\end{abstract}
\noindent{\it Keywords\/}: Special issue, chemical kinetics, stochastic
processes, population dynamics
%
%
%
\section{Introduction}
Pain in animals, including humans, is triggered by the so-called
nociceptors, sensory neurons that react to potentially damaging
stimulus. Neurology textbooks \cite{VR2001} reports on the cascade of successive
reactions which are activated by the so called noxious stimuli: The
peripheral terminals of primary sensory neurons launch the signal,
which is then transmitted to the spinal and supraspinal nuclei and
eventually yields to the activation of a matrix of cortical areas
that are deputed to the conscious experience of pain.

More specifically, the stimulus originating from a bodily harming
menace can be directly processed through transduction by the
receptors located on the nerve terminals. Alternatively, an indirect
pathway can take over through the activation of transient receptor
potentials on keratinocytes or the release of intermediate molecules
(such as the ATP) which, in turn, act on sensory neurons receptors.
In the following we shall assume the first scenario to hold, and,
though certainly important, disregard other mechanisms that might be
simultaneously in play. In other words, we simplistically imagine
that pain receptors act as effective gates, channeling the route to
the involved cortical circuits.

Analgesic drugs relieve  the pain by interfering with the peripheral
and central nervous system. Drug molecules bind in fact their target
receptors, and consequently inhibit the pain perception. To grasp
and visualize the essence of the process, one can hypothesize that
the bound chemical element occludes the path, by impeding the signal
transduction through the channel envisioned above.

Analgesic are commonly used in basic research and clinical practice,
but their interaction with nociceptory and normal sensory processing
remains to be fully unraveled. Anesthetics are for instance known to
modify the electrical recordings measured via evoked potentials
(EPs) responses \cite{MF2001}, a powerful diagnostic tools employed to monitor
and
characterize a large variety of central nervous system disorders.
EPs are elicited by a specific stimulus applied to the e.g. pain
receptors and consist in recording the induced electrical brain
activity, as detected by localized electrodes placed on the surface
of the head. Furthermore, EPs are also useful in documenting
objective response to pain \cite{DC2005,GSTP2007} and can thus prove
fundamental to
elucidate the molecular processes that control anesthetic absorption
and metabolization.

Different analgesic agents have been shown to produce intriguingly
distinct effects at the level of the EPs \cite{RNR2006}. Recorded
time series of the solicited electric activity display in fact
remarkably different patterns, which are generically attributed to
the chemical specificity of the anesthetic compound. Qualitatively,
large, regular, oscillations of the electric response manifest,
latency and amplitude being peculiar traits, supposedly related to
the molecular characteristic of the administered drug.

Furthermore, cycles in the perception of pain have been also
reported which might be hypothetically driven by similar microscopic
processes, the  interaction between the anaesthetic molecules and
their targets playing certainly a role of paramount importance.
Clearly, the individual experience of pain is also influenced by
psychological and cultural factors, unfortunately difficult to
deconvolve when aiming at resolving the objective
picture.

The issue of developing a unique  interpretative framework to
account for the presence of such oscillatory regimes has catalyzed
vigorous discussions. The puzzle of their existence remains however
to be fully understood.

Current mathematical models \cite{DJDDPV2007} approach  the problem via
deterministic paradigms, thus neglecting the crucial role which is certainly
played by the noise, intrinsic to the phenomenon under scrutiny.
These aspects become particularly important when accounting for the
presence of diverse chemical species, which populate the stream flow
in a spatially diffusive environment. Different chemical entities
may compete with the drug molecules and occupy the sites located in
close vicinity of the receptors, thus effectively hindering the
binding event. Under specific conditions, such competition sustained
by the stochastic component of the dynamics might result in large
temporal oscillations for the amount of bound receptors, a mechanism
which could explain the emergence of macroscopic cycles for the
sensation of pain in response to medicaments.

In this paper, we shall speculate  on the above scenario by putting
forward a network of chemical reactions and performing a system-size
expansion through the celebrated van Kampen theory
\cite{vanKampen1992}. This enables us to derive a set of linear
equations for the fluctuations, with coefficients related to the
steady-state concentrations predicted from the first-order theory
(i.e. the deterministic rate equations). Solutions are identified
for which the deterministic steady-state occurs via damped
oscillations: the inclusion of second-order fluctuations leads then
to the amplification of sustained oscillations. These conclusions
are briefly discussed with reference to the existing medical
literature.
\section{Description of the model}
Within the simplified scenario depicted  above, we shall model the
chemical interaction between a large, though finite, number of drug
molecules (anesthetic), hereby termed $T$, and free receptors $R_F$
which represent their binding target. Following a successful binding
event, a molecule of the species $T$ disappears, leaving an empty
case, hereafter labelled $E$. The population of bound receptor $R_T$
is in turn increased by one unit. These assumptions formally
translate into the compact chemical notation:
\begin{equation}
\label{chem1}
 R_F + T  \stackrel{\alpha}{\longrightarrow}  R_T +E
\end{equation}
where $\alpha$ stands for the  associated reaction rate. The inverse
reaction corresponding to the spontaneous detachment of the bound
component may occur \footnote{We here assume that the free $T$
molecule is still chemically active and can thus potentially chase
for unscreened targets. This working hypothesis can be relaxed
leading to conclusions qualitatively similar to the ones highlighted
below.} with a certain probability $\beta$\footnote{In principle it
would be extremely useful to dispose of experimental  estimates for the
reaction rates, so to define a realistic range of variability for the free
parameters in the model. The most reliable data concern the
so-called (equilibrium) affinity constant for the case of e.g. the Tramadol, an
analgesic agent which belongs to the class of synthetic opioid. Depending on the
target receptor (and on the specificity of the chaser's molecule) the affinity
constant is reported to vary of a large amount which scans two orders of
magnitude (from fraction of unity to hundreds) \cite{GS2004,GHKW2000}.}, which
motivates the
introduction of the dual relation:
\begin{equation*}
 R_T + E  \stackrel{\beta}{\longrightarrow} R_F +T
\end{equation*}

\begin{figure}[t]
\begin{tabular}{|lcr|}
  \hline
  \phantom{a} & \phantom{a} & \phantom{a} \\
  \begin{minipage}[h]{3 cm}
  \begin{eqnarray*}
  R_F + T  & \stackrel{\alpha}{\longrightarrow} &  R_T +E\\
  R_T + E & \stackrel{\beta}{\longrightarrow}  & R_F +T
  \end{eqnarray*}
  \end{minipage}  & \phantom{space space space space} &
\begin{minipage}[h]{6cm}
\includegraphics[width=5 cm,height=4 cm]{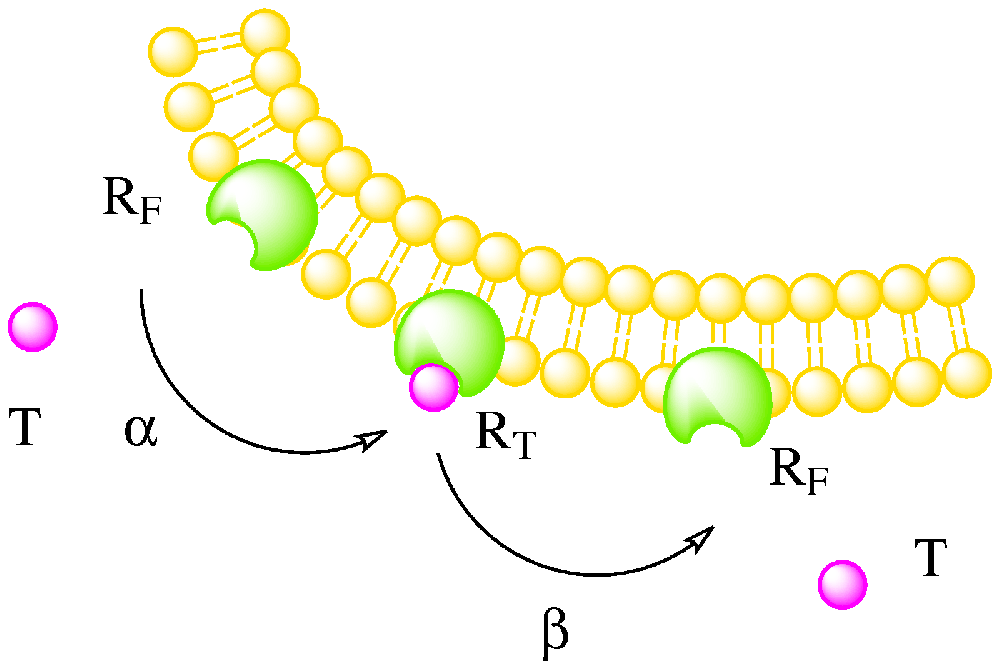}
  \end{minipage}  \\
   \phantom{a} & \phantom{a} & \phantom{a} \\
  \hline
   \phantom{a} & \phantom{a} & \phantom{a} \\
  \begin{minipage}[h]{3 cm}
  \begin{eqnarray*}
  H + T   & \stackrel{\gamma}{\longrightarrow} &H + H  \\
  H + T   & \stackrel{\sigma}{\longrightarrow}& H + E
  \end{eqnarray*}
  \end{minipage}  & \phantom{space space space space} &
\begin{minipage}[h]{6cm}
\includegraphics[width=5 cm,height=5 cm]{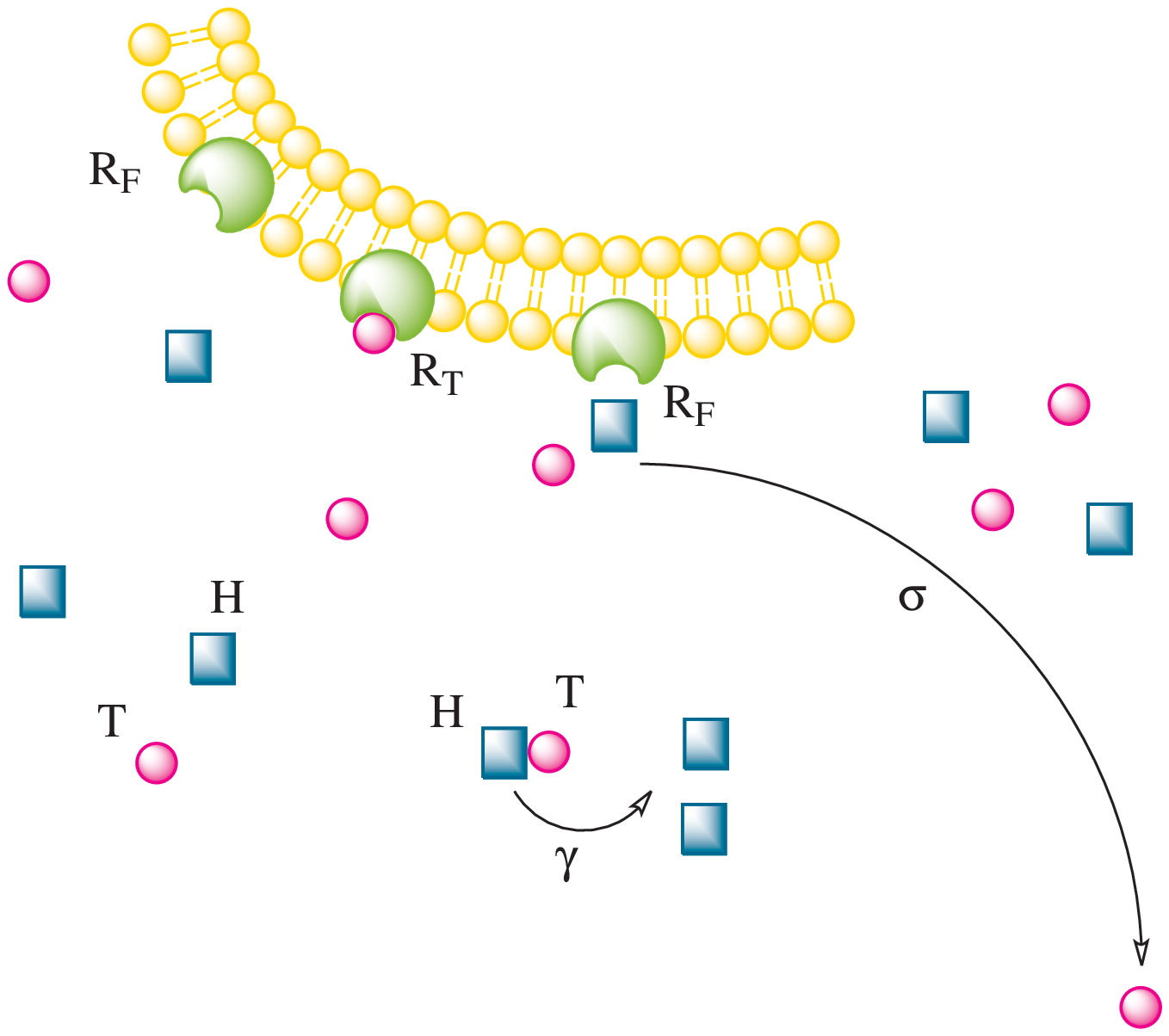}
  \end{minipage}  \\
  \phantom{a} & \phantom{a} & \phantom{a} \\
  \hline
  \end{tabular}
  \caption{The main reaction schemes are depicted. The squares stand for the inactive species, while the circles
represent the drug molecules. The model is then complemented with a
set of additional reactions (see equations
(\ref{chemT})--(\ref{chemH})), which accounts for the possibility
that $T$ and $H$ enter (resp. leave) the region deputed for the
interaction. }\label{fig_chemical_equations}
\end{figure}

The anesthetic molecules $T$ surf in a densely packed environment
and certainly experience collisions with several other microscopic
chemical entities, $H$, which populate the streaming flow. Binary
interactions between $H$ and $T$ elements, can occur in the close
vicinity of the receptors ($R_F$) location, potentially disturbing
and eventually interfering with the binding event. As a result of an
hypothetical collision, the active species $T$ can be ejected by the
spatial layer immediately adjacent to the receptor, leaving behind
an empty case $E$. Beyond this effect, which stems from purely
steric interactions, one has to account for possible chemical
transformations,  which might occur when individuals from the $H$
and $T$ species encounter: The active chaser $T$ can lose its
ability to bind the target \footnote{Note that this can happen both
due to a mechanical stress or via chemical combination of the
colliding species, see for instance \cite{Katzung2004} where the plasma
protein binding is discussed. For a specific application relative to the case
of the  Tramadol refer to \cite{LBOS1986}.} and it is
thus mapped into an inactive $H$ molecule. To incorporate these
effects into the proposed description we postulate the following
interaction rules, which are loosely inspired to the predator-prey
competition mechanism:
\begin{eqnarray*}
 H + T   & \stackrel{\gamma}{\longrightarrow} &H + H  \\
 H + T   & \stackrel{\sigma}{\longrightarrow}& H + E
\end{eqnarray*}
The values of $\sigma$ and $\gamma$  characterize the effectiveness
of the interaction, which is in turn sensitive to the choice of the
compound $T$. The idealized cartoons of figure
\ref{fig_chemical_equations} are aimed at visualizing the above
reaction schemes.

To complete the model we introduce an effective  migration, by
requiring that the $T$ and $H$ molecules can enter (resp. leave) the
region deputed to the interaction. The latter assumption yields to
the following set of chemical relations:
\begin{eqnarray}
\label{chemT} T       & \stackrel{\delta_1}{\longrightarrow}& E  \\
 E       & \stackrel{\eta_1}{\longrightarrow} &T  \\
 H       & \stackrel{\delta_2}{\longrightarrow}& E  \\
\label{chemH} E       & \stackrel{\eta_2}{\longrightarrow} &H
\end{eqnarray}
The population, namely the ensemble of elements belonging to an
homologous species $X$, will be labelled in the following with the
symbol $n_X$. Notice that the number of receptors  $N_1 =
n_{R_T}+n_{R_F}$ and the total amount of molecules (including the
empties) $N_2=n_T + n_H + n_E$ are conserved quantities. This
observation enables us to reduce the complexity of the problem by
setting:
\begin{equation*}
n_{R_F}=N_1-n_{R_T} \qquad n_E = N_2-n_T - n_H
\end{equation*}
In the following we shall use the vectorial notation $\underline{n}
=(n_T,n_{R_T},n_H)$ to help keeping  the mathematical developments
compact.

We are now in a position to define the transition rates
$T(\underline{n}^{'}|\underline{n})$ from a state $\underline{n}$ to
a different one $\underline{n}^{'}$. In our convention initial
states are on the right and final states on the left. As an example,
consider  equation (\ref{chem1}). The probability to pick up a $T$
constituents follows from simple combinatorics and reads $n_T / N_2$, while
there is a probability
$(N_1-n_{R_T})/N_1$ of $R_F$ being chosen. This results in $\alpha
(n_T / N_2) (N_1-n_{R_T})/N_1$ for this particular transition rate
\cite{MN2004}.
A complete listing of transition probability associated to the
preceding set of chemical reactions is here enclosed:
\begin{eqnarray*}
T(n_T-1,n_{R_T}+1,n_H|\underline{n}) & = & \alpha
\frac{n_T}{N_2}\frac{N_1-n_{R_T}}{N_1} \\
T(n_T+1,n_{R_T}-1,n_H|\underline{n}) & = & \beta    \frac{N_2 - n_T
-n_H}{N_2}\frac{n_{R_T}}{N_1} \\
T(n_T-1,n_{R_T},n_H+1|\underline{n}) & = & \gamma
\frac{n_H}{N_2}\frac{n_T}{N_2} \\
T(n_T-1,n_{R_T},n_H|\underline{n})   & = & \sigma   \frac{n_H}{N_2}
\frac{n_T}{N_2} +\delta_1 \frac{n_T}{N_2} \\
T(n_T+1,n_{R_T},n_H|\underline{n})   & = & \eta_1   \frac{N_2 - n_T
-n_H}{N_2}\\
T(n_T,n_{R_T},n_H-1|\underline{n})   & = & \delta_2 \frac{n_H}{N_2}\\
T(n_T,n_{R_T},n_H+1|\underline{n})   & = & \eta_2   \frac{N_2 - n_T -n_H}{N_2}
\end{eqnarray*}

Having defined the transition rates, the master equation governing
the evolution of the discrete stochastic model takes the form:
\begin{eqnarray}\label{eq:masterEquation}
 \frac{\mathrm{d}}{\mathrm{d} t} P(\underline{n},t) & = &
 T(\underline{n}|n_T+1,n_{R_T}-1,n_H)P(n_T+1,n_{R_T}-1,n_H,t)       \nonumber\\
 & + & T(\underline{n}|n_T-1,n_{R_T}+1,n_H)P(n_T-1,n_{R_T}+1,n_H,t) \nonumber\\
 & + & T(\underline{n}|n_T+1,n_{R_T},n_H-1)  P(n_T+1,n_{R_T},n_H-1,t)
\nonumber\\
 & + & T(\underline{n}|n_T+1,n_{R_T},n_H)P(n_T+1,n_{R_T},n_H,t) \nonumber\\
 & + & T(\underline{n} |n_T-1,n_{R_T},n_H) P(n_T-1,n_{R_T},n_H,t)   \nonumber\\
 & + & T(\underline{n} |n_T,n_{R_T},n_H+1) P(n_T,n_{R_T},n_H+1,t)   \nonumber\\
 & + & T(\underline{n} |n_T,n_{R_T},n_H-1) P(n_T,n_{R_T},n_H-1,t)   \nonumber\\
 & - & \Big[ T(n_T-1,n_{R_T}+1,n_H|\underline{n})
       +T(n_T+1,n_{R_T}-1,n_H|\underline{n})                        \nonumber\\
 & + & T(n_T-1,n_{R_T},n_H+1|\underline{n})
       +T(n_T-1,n_{R_T},n_H|\underline{n})                        \nonumber\\
& + & T(n_T+1,n_{R_T},n_H|\underline{n})
       +T(n_T,n_{R_T},n_H-1|\underline{n})                        \nonumber\\
 & + & T(n_T,n_{R_T},n_H+1|\underline{n}) \Big] P(\underline{n},t)
\end{eqnarray}
where $P(\underline{n},t)$ is the probability  to find the system in
the state $\underline{n}$ at time $t$. In the next section we shall
shortly report about our stochastic simulations, before turning to
develop the analytical framework.
\section{Numerical simulations}
Based on the chemical equations introduced above, numerical
simulations can be carried on, which respect the  intrinsic
stochastic nature of the model. To this end we employ the celebrated
Gillespie algorithm \cite{Gillespie1976}  which exploits the
information encoded in the reaction scheme to advance the system in
time through a sequence of random increments\footnote{The standard
implementation of the Gillespie algorithm is based on a nested sequence of
operations which is here briefly recalled. First, one initializes the system at
$t
=0$, by assigning a number of
molecules to each of the considered species. Then
the following iterative
scheme runs: (i) Calculate the
transition rates $T_i(\underline{n}'|\underline{n})$
associated to each prescribed reaction $i$. The quantity  $T_0 =  \sum_{i=1}^M
T_i(\underline{n}'|\underline{n})$ is stored; (ii) Extract two random numbers,
respectively $r_1$ and $r_2$,  from a uniform distribution, which are used to
(a) update the simulation time by a finite amount
$\delta t = 1/T_0 \ln(1/r_1)$ and (b) select the index $i$ labelling the next
reaction which is deputed to occur ($i$ is such that $\sum_{k=1}^{i-1} T_k < r_2
T_0 \leqslant \sum_{k=1}^{i} T_k$); (iii) Update the species accordingly and
go back to point (i).}. A randomly selected reaction is forced to occur during
each
successive step. It should
be emphasized that time increments and associated reactions are
chosen so to recover the exact probability distribution of the
stochastic time series. For a more detail account on the philosophy
of the integration recipe, the reader can consult
\cite{Gillespie1976}.

\begin{figure}[tb]
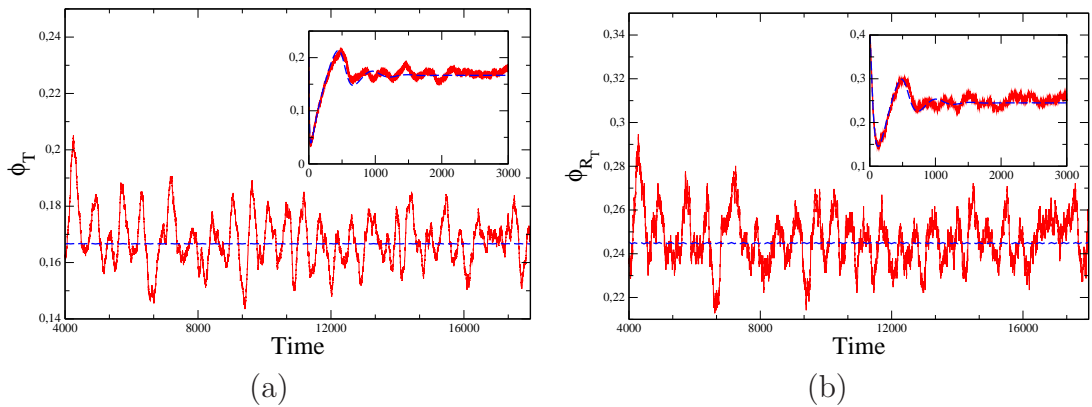

\begin{center}
\vspace*{2.5em}
\begin{tabular}{cc}
  \includegraphics[width=7cm]{figure2a.eps} &
  \includegraphics[width=7cm]{figure2b.eps} \\
(a) & (b)
\end{tabular}
\caption{Drug molecules (panel (a)) and  bound receptor (panel (b)) densities
as a function of time. The solid lines refer to the time series from
stochastic simulations, while the dashed lines are calculated from
numerical integration of the mean field equations
(\ref{meanfield1})-(\ref{meanfield3}). The insets represent a zoom of the
initial evolution and allow to appreciate the agreement with the
mean--field solution at short times. Parameters values
for both
panels are $\alpha=0.008$, $\beta=0.005$, $\gamma=0.3$,
$\delta_1=0.001$, $\delta_2=0.05$, $\eta_1=0.001$, $\sigma=0.06$,
$N_1=5300$ and $N_2=20000$.} \label{fig1}
\end{center}
\end{figure}

A typical result is represented  in figure \ref{fig1} where the
normalized population of $T$-like molecules and bound receptors
$R_T$ are reported, as function of time. Notice that large
stochastic oscillations are clearly displayed, despite the
relatively large number of simulated molecules. Even more
interestingly, the oscillations persist when increasing the
populations amount and only when a very large number of constituents
is allowed, the system eventually sets down to its mean-field
solution.

As already remarked  in \cite{MN2004,MN2005,MNNS2007}, this result
is odds with the intuitive believe that fluctuations can be safely
ignored, due to the usual statistical factor $1/\sqrt{N}$. Indeed,
the observed fluctuations arise from the amplification of the
intrinsic noise, associated to the discrete component of the
dynamics and can potentially bear and extraordinary conceptual
relevance in several applications. With reference to the case at
hand, emerging, regular, oscillatory patterns in the $R_T$ quota,
could potentially explain the modulations in the pain perception as
reported in the literature, higher values of $n_{R_T}$
corresponding, in our interpretative scheme, to a less pronounced
pain level. Alternatively, on a different time scale, the effects
here discussed could provide a consistent microscopic picture to
understand the presence of quasi--periodic fluctuations in the evoked
electric activity of laboratory animals under anesthetic treatment.

In the forthcoming sections, we shall gain some analytical insight
into the model results and characterize the specific traits of the
observed oscillatory behaviors through the elegant van Kampen
\cite{vanKampen1992} expansion.
\section{On the deterministic limit}
The deterministic counterpart of the governing master equation is
straightforwardly obtained as follows. Focus on $T$ and observe that, by
definition:
\begin{equation*}
\left\langle n_T \right\rangle = \sum_{\underline{n}} n_T P(\underline{n},t)
\end{equation*}

Multiplying equation (\ref{eq:masterEquation}) by $n_T$ and summing
over $\underline{n}$ returns on the right--hand side $d \left\langle
n_T \right\rangle / d t$. Simplifying  the left--hand side is
somehow more laborious and requires some effort. Proceeding in a
completely analogous fashion for $n_H$ and $n_{R_T}$ yields  to the
following system of coupled differential equation for the ensemble
independent variables:
\begin{eqnarray*}
\fl\frac{\mathrm{d}}{\mathrm{d} t} \langle n_T \rangle = &
-\alpha \left\langle \frac{n_T}{N_2}\frac{N_1-n_{R_T}}{N_1} \right\rangle +\beta
\left\langle \frac{N_2 - n_T -n_H}{N_2}\frac{n_{R_T}}{N_1}  \right\rangle -(
\gamma +\sigma) \left \langle \frac{n_H}{N_2}\frac{n_T}{N_2} \right\rangle  \\
\fl &-\delta_1 \left\langle \frac{n_T}{N_2}\right \rangle +\eta_1
\left\langle \frac{N_2 - n_T -n_H}{N_2}\right \rangle \\
\fl\frac{\mathrm{d}}{\mathrm{d} t} \langle n_{R_T} \rangle  = &
\alpha \left\langle \frac{n_T}{N_2}\frac{N_1-n_{R_T}}{N_1} \right\rangle -\beta
\left \langle \frac{N_2 - n_T -n_H}{N_2}\frac{n_{R_T}}{N_1} \right \rangle \\
\fl\frac{\mathrm{d}}{\mathrm{d} t} \langle n_H \rangle  = & \gamma \left\langle
\frac{n_H}{N_2}\frac{n_T}{N_2}  \right\rangle +\eta_2 \left\langle  \frac{N_2 -
n_T -n_H}{N_2}\right\rangle -\delta_2 \left\langle \frac{n_H}{N_2}\right\rangle
\end{eqnarray*}

The mean--field approximation corresponds to ignore the
correlations, in the above rate equations when performing the limit
for $N_1$ and $N_2$ large. Introducing $\phi_T=\langle n_T\rangle/N_2$,
$\phi_{R_T}=\langle n_{R_T}\rangle/N_1$, $\phi_H=\langle n_H\rangle/N_2$,
rescaling time as $\tau= t/ N_2$ and formally sending $N_1,N_2 \rightarrow
\infty$, one finally obtains:
\begin{eqnarray}
\label{meanfield1}\frac{\mathrm{d}}{\mathrm{d}\tau}\phi_T =
&-\alpha\phi_T(1-\phi_{R_T}) +
\beta\phi_{R_T} (1-\phi_T- \phi_H) -(\gamma+\sigma) \phi_H \phi_T \\
&-\delta_1 \phi_T+\eta_1(1-\phi_T- \phi_H) \nonumber\\
\label{meanfield2}\frac{\mathrm{d}}{\mathrm{d}\tau}\phi_{R_T} = & c\left[ \alpha
\phi_T(1-\phi_{R_T}) - \beta \phi_{R_T} (1-\phi_T- \phi_H) \right]  \\
\label{meanfield3}\frac{\mathrm{d}}{\mathrm{d}\tau}\phi_H = & \gamma \phi_H
\phi_T + \eta_2(1-\phi_T- \phi_H)-\delta_2 \phi_H
\end{eqnarray}
where $c=N_2/N_1$.

We shall hereon limit our discussion to the case $\eta_2=0$ which
will prove analytically tractable. The conclusions here demonstrated
with reference to the selected case study, will obviously apply to
the more general setting where fresh $H$ molecules are allowed to
enter the interaction region. Investigations on the complete model
($\eta_2 \ne 0$) will be object of a forthcoming publication.

Two fixed points, respectively labelled  $\underline{\phi}_1^*$ and
$\underline{\phi}_2^*$, are identified:
\begin{eqnarray*}
\fl& \underline{\phi}_1^*=\left ( \frac{\eta_1}{\delta_1 + \eta_1} ,
\frac{\alpha \eta_1}{\beta \delta_1 + \alpha \eta_1} , 0\right) \\
\fl& \underline{\phi}_2^*=\left ( \frac{\delta_2}{\gamma}, \frac{\alpha
[\delta_2(\gamma+\sigma)+\eta_1 \gamma]}{\alpha [\delta_2(\gamma+\sigma)+\eta_1
\gamma]+\beta [(\gamma+\sigma)(\gamma -\delta_2)+\delta_1 \gamma]},
\frac{\eta_1(\gamma -\delta_2)-\delta_1 \delta_2}{\delta_2(\gamma + \sigma) +
\eta_1 \gamma}\right)
\end{eqnarray*}

In figure \ref{fig1}, the solid lines represent the above
equilibrium solution: For such choice of the parameter, as
previously mentioned, the stochastic dynamic displays macroscopic
oscillation for the monitored quantities, the average reference
value being correctly predicted by the mean-field theory. What is
the reason for these regular fluctuation to manifest? Are they
resulting from the intrinsic finite size nature of the simulated
medium?

To answer these questions it is crucial to determine the stability
matrix associated with the fixed points, as it will play a central
role in investigating the cycling phenomenon. One can thus re--write
the mean--field equations in the compact form $d \phi_k / d \tau =
f_k(\underline{\phi})$, where the index $k = 1,2,3$  codes the
different species, namely $T$ ($k=1$), $R_T$ ($k=2$) and $H$
($k=3$).  The $3 \times 3$ Jacobian matrix $J_{ij} = \partial f_i /
\partial \phi_j |_{FP}$ (here FP means ``evaluated at the fixed
point'') controls the linearized dynamics about the fixed point and
can be cast in the form:

\begin{tiny}
\begin{equation}
\label{Jac} \fl J(\underline{\phi}^*) = \left (
\begin{array}{ccc}
-\alpha(1-\phi_{R_T}^*) -\beta \phi_{R_T}^* -(\gamma+\sigma) \phi_H^*
-\delta_1
-\eta_1  & \alpha \phi_T^*+\beta(1 -\phi_T^* -\phi_H^*) & -\beta \phi_{R_T}^*
-(\gamma+\sigma)\phi_T^* -\eta_1 \\
c[\alpha(1-\phi_{R_T}^*) +\beta\phi_{R_T}^*]  & -c[\alpha \phi_T^*+\beta(1
-\phi_T^* -\phi_H^*)] & c\beta \phi_{R_T}^*\\
\gamma \phi_H^* & 0 & \gamma \phi_T^* -\delta_2
 \end{array}
 \right)
\end{equation}
\end{tiny}

One can easily show that $J(\phi^*_1)$ admits two real negative
eigenvalues and a third one, also real, whose sign depends on the
choice of the parameters. A stability analysis for the second
equilibrium point proves technically difficult. However, via
numerical inspections, a large region in the parameters' space is
identified, which yields to complex solutions. In particular,
complex eigenvalues of the $J(\phi^*_2)$ are found having negative
real part. This condition ensures an oscillatory approach to
equilibrium, a fundamental ingredient which is eventually
responsible for the large scale modulations observed in the finite
size regime. Tracing the region in space deputed to the
aforementioned behavior falls out of the scope of the present paper
and shall be postponed to a forthcoming contribution together with a
detailed characterization of the general case with $\eta_2 \neq 0$.

Following the above, from now on, we shall denote with $\underline\phi^*$ a
particular value of $\underline\phi^*_2$ for which damped oscillations do occur
in the mean--field scenario.
\section{Characterizing the fluctuations: The van Kampen expansion}
As clearly depicted  in figure \ref{fig1} the innate discreteness of
the stochastic medium drives into the system important effects which
cannot be captured within the continuous mean--field scenario. To
shed light into the observed phenomena we can bring into the game
the fluctuations by performing the following explicit replacement:
\begin{equation}
\label{xi}
\fl n_T= N_2 \phi_T(t) +\sqrt{N_2} \xi_T  \quad n_{R_T}= N_1 \phi_{R_T}(t)
+\sqrt{N_1} \xi_{R_T} \quad  n_H= N_2 \phi_H(t) +\sqrt{N_2} \xi_H
\end{equation}
where the new continuous  variable $\underline{\xi}=(\xi_T, \xi_{R_T}
\xi_{R_H})$ replace the discrete quantity $\underline{n}=(n_T,
n_{R_T}, n_H)$  in the definition the probability distribution,
namely $P(\underline{n},t)=\Pi(\underline{\xi},\tau)$.

Before proceeding, it is worth  emphasizing that the $1/\sqrt{N_1}$
(resp. $1/\sqrt{N_2}$) term holds because of the central-limit
theorem: The fluctuations are in fact expected to decay in a similar
fashion and, in the continuous limit, $N_1,N_2 \rightarrow \infty$
the system is entirely characterized in term of its continuous
variables $\underline{\phi}=(\phi_T, \phi_{R_T}, \phi_H)$  as
prescribed by equations (\ref{xi}).

The master equation can be re-written as function of the new
variables. The left-hand side reads:
\begin{equation*}
\fl\frac{\mathrm{d}}{\mathrm{d} t} P(\underline{n},t)
=\frac{1}{N_2}\frac{\partial \Pi}{\partial
\tau}-\frac{1}{\sqrt{N_2}}\frac{\mathrm{d}}{\mathrm{d} \tau} \phi_T
\frac{\partial \Pi}{\partial
\xi_T}- \frac{c^{-1}}{\sqrt{N_1}}\frac{\mathrm{d}}{\mathrm{d} \tau} \phi_{R_T}
\frac{\partial \Pi}{\partial \xi_{R_T}}
-\frac{1}{\sqrt{N_2}}\frac{\mathrm{d}}{\mathrm{d} \tau} \phi_H
\frac{\partial \Pi}{\partial \xi_H}
\end{equation*}
where use has been made of the fact that the time derivative is
taken at constant $\underline{n}$. The right-hand side follows from
a straightforward, though lengthy, application of the large-$N$ van
Kampen expansion. The main step of the derivation are reviewed in the
appendix. The interested reader can refer to
\cite{vanKampen1992, Gardiner1985} for a detailed account on the
whole procedure.

At leading order of the expansion we recover the deterministic mean--field
equations (\ref{meanfield1})--(\ref{meanfield3}),
while at next--to--leading order we obtain the linear multivariate Fokker
Planck
equation (\ref{FP}) that governs the evolution of the fluctuations. The
coefficients of
this equation are completely specified as function
of the model's parameters (see the appendix): In principle, by solving
equation (\ref{FP}) we are in a position
to quantify the deviation from the
ideal mean--field dynamics, via the probability distribution $\Pi$.
At present, we aim at understanding the oscillation and, to this
end, we invoke a completely equivalent formulation of the Fokker Planck
equation. The problem can be in fact
cast as a set of stochastic differential equations of
Langevin type, which take the explicit form:
\begin{equation}
\label{langevin}
\frac{d \xi_i}{d \tau} = A_i(\underline{\xi}) + \eta_i (\tau)  \qquad i=1,..,3
\end{equation}
where for convenience, as a natural extension  of the notation
introduce above, we have now set $\xi_1=\xi_T$, $\xi_2=\xi_{R_T} $,
$\xi_3=\xi_{H}$ and $A_i$ is specified in the appendix. The term $\eta_i$ is a
Gaussian noise with zero mean and with correlation given by  
\begin{equation*}
\langle \eta_i (\tau) \eta_j (\tau^{'})\rangle = B_{ij} \delta (\tau-\tau^{'})
\end{equation*}

\begin{figure}[htbp]
  \centering
  \vspace*{2.5em}
  \includegraphics[width=7cm]{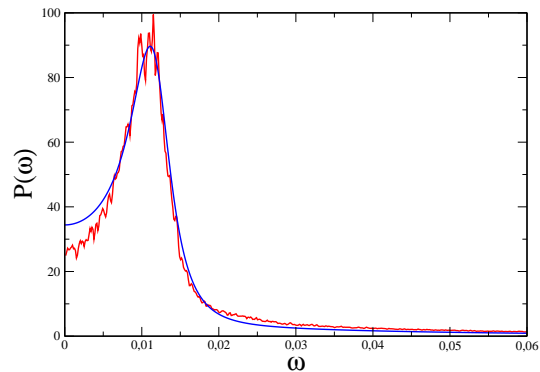}
  \caption{A plot of the power  spectrum $P(\omega)$  for the $R_T$ time series
as a function of the frequency $\omega$. The noisy  line corresponds
to the spectrum calculated from 500 runs of the stochastic simulation
of the model. The smooth line shows the analytic prediction from
equation (\ref{ps}). For the parameters' setting refer to the
caption of figure
  \ref{fig1}.}
  \label{fig2}
\end{figure}

To highlight the existence of a possible  oscillatory behavior we
perform a Fourier analysis of equations (\ref{langevin}), and
obtain:
\begin{equation*}
-i \omega \tilde{\xi}_i (\omega) = \sum_j M_{ij} \tilde{\xi}_j(\omega)+
\tilde{\eta}_i (\omega)
\end{equation*}
where the tilde stands for the Fourier transform. Following
\cite{MNNS2007} we can re-write this as:
\begin{equation}
\label{fourier}
\sum_j \Phi_{ij} (\omega) \tilde{\xi}_j(\omega) = \tilde{\eta}_i(\omega)
\end{equation}
with $\Phi_{ij} (\omega) = -i \omega \delta_{ij}-M_{ij}$. In addition one gets:
\begin{equation*}
\langle \tilde{\eta}_i (\omega) \tilde{\eta}_j (\omega^{'})\rangle = B_{ij}(2
\pi)
\delta (\omega+\omega^{'})
\end{equation*}
Solving equation (\ref{fourier}) for $\tilde{\xi}_i$ and  computing
the power spectrum results in:
\begin{equation}
\label{ps}
P_i (\omega) = \langle |\tilde{\xi}_i(\omega)|^2\rangle = \sum_j \sum_k \Phi_{ij}^{-1}(\omega) B_{jk}
(\Phi_{ij}^{\dagger})^{-1}(\omega)
\end{equation}
where we have  used $\Phi_{ij}^{\dagger}(\omega)=\Phi_{ji}(-\omega)$. The power
spectrum predicted by
equation (\ref{ps}) is plotted in figure \ref{fig2}, for the same
set of parameters as employed in the simulations of  figure
\ref{fig1}. A clear peak is detected. Moreover, the theoretical
curve interpolates correctly the numerical profile. These results
confirm that the macroscopic oscillations which manifest in the
acquired time-series stem from the noise
intrinsic to the system under investigation. It is our believe that
mechanisms similar to the ones here hypothesized, are potentially in
place in the complex human (animal) body environment and  could in
principle form the basis of a consistent molecular interpretation
for the large collection of experimental, biomedical observations to
which we made reference in the introductory section.
%
%
%
\section{Conclusions}
In this paper we propose a discrete dynamical framework which is
aimed at shedding new light onto the complex molecular processes
intervening in response to an external harming stimulus  so to
trigger the pain sensation. We are in particular interested in
elucidating the crucial interplay between the administered drug
molecules, which express their analgesic function chasing the target
receptors, and  other chemical elements freely diffusing in the
stream. The latter can substantially reduce the anaesthetic effect,
by hindering the available binding sites. Similarly, drug molecules
can be turned into inactive species following binary encounters. The
mechanisms here postulated are formally coded via chemical reactions
and define a consistent stochastic scheme. Numerical simulations
display macroscopic oscillations in the concentration amount: the
number of bound receptors change cyclically in time, a trend which
we assume to induce an analogous modulation for the experienced
perception of pain. These findings are analytically illustrated by
developing a large-size expansion which enables us to predict the
existence of a peaked power spectrum. It is important to remark that
the  amplification process here discussed stems from the underlying
stochasticity, which excites the resonant frequencies of the system.
Oscillations arise hence naturally, driven by the noise which is
intrinsic to the system and without invoking any ad hoc couplings
among the molecular agents participating to the  dynamics. Our
findings could suggest the existence of a simple, though general,
molecular mechanism responsible for the emergence of cyclic
behaviors in response to analgesic treatments \cite{RNR2006}. We shall here
stress that our main conclusions apply also for the more general setting where
$\eta_2 \neq 0$. In particular the peaked power spectrum is also found in this
latter case and
the region in the parameters' space which corresponds to the emergence of the
cycles can be partially identified on the basis of explicit
analytic formulae. These findings will be reported in a forthcoming publication
\cite{DF2008}.
\appendix
\setcounter{section}{1}
\section*{Appendix}
The first technical point of the van Kampen expansion concerns the  introduction
of the
so--called shift operators, $\mathbb{E}^{\pm 1}_T, \mathbb{E}^{\pm
1}_{R_T}, \mathbb{E}^{\pm 1}_H$ which obey:
\begin{eqnarray*}
 & \mathbb{E}^{\pm 1}_T  f(\underline{n},t) = f(n_T \pm 1 , n_{RT}, n_H) \\
 & \mathbb{E}^{\pm 1}_{R_T}  f(\underline{n},t)= f(n_T, n_{RT} \pm 1 , n_H) \\
 & \mathbb{E}^{\pm 1}_H f(\underline{n},t) = f(n_T , n_{RT}, n_H \pm 1) \: .
\end{eqnarray*}
The master equation (\ref{eq:masterEquation}) is hence cast in the
form:
\begin{eqnarray}
\label{master_mod}
\fl \frac{1}{N_2}\frac{\partial \Pi}{\partial
\tau} &-&\frac{1}{\sqrt{N_2}}\frac{\mathrm{d}}{\mathrm{d} \tau} \phi_T
\frac{\partial \Pi}{\partial
\xi_T} - \frac{c^{-1}}{\sqrt{N_1}}\frac{\mathrm{d}}{\mathrm{d} \tau}
\phi_{R_T}
\frac{\partial \Pi}{\partial \xi_{R_T}}
-\frac{1}{\sqrt{N_2}}\frac{\mathrm{d}}{\mathrm{d} \tau} \phi_H
\frac{\partial \Pi}{\partial \xi_H} =  \nonumber \\
 \fl& + & (\mathbb{E}^{+1}_T  \mathbb{E}^{-1}_{R_T} -1) \left[ \alpha
\left( 1-\phi_{R_T}-\frac{1}{\sqrt{N_1}} \xi_{R_T} \right ) \left(\phi_T
+\frac{1}{\sqrt{N_2}} \xi_T \right ) \Pi \right ] \nonumber\\
\fl& +  & (\mathbb{E}^{-1}_T \mathbb{E}^{+1}_{R_T} -1) \left[ \beta \left(
\phi_{R_T}+\frac{1}{\sqrt{N_1}} \xi_{R_T} \right ) \left( 1 -\phi_T -
\frac{1}{\sqrt{N_2}} \xi_T -\phi_H -\frac{1}{\sqrt{N_2}} \xi_H \right ) \Pi
\right ] \nonumber\\
\fl& +  & (\mathbb{E}^{+1}_T \mathbb{E}^{-1}_H -1) \left[ \gamma \left( \phi_T +
\frac{1}{\sqrt{N_2}} \xi_T\right )\left( \phi_H
+\frac{1}{\sqrt{N_2}}\xi_H \right ) \Pi\right ] \nonumber\\
\fl& +  & (\mathbb{E}^{+1}_T  -1) \left[ \sigma \left( \phi_T +
\frac{1}{\sqrt{N_2}} \xi_T\right )\left( \phi_H
+\frac{1}{\sqrt{N_2}}\xi_H \right ) \Pi  +\delta_1  \left( \phi_T +
\frac{1}{\sqrt{N_2}} \xi_T\right )\Pi \right ]\nonumber \\
\fl& +  & (\mathbb{E}^{-1}_T  -1) \left[ \eta_1 \left( 1 -\phi_T -
\frac{1}{\sqrt{N_2}} \xi_T -\phi_H -\frac{1}{\sqrt{N_2}} \xi_H \right ) \Pi
\right ]\nonumber \\
\fl& +  & (\mathbb{E}^{+1}_H  -1) \left[ \delta_2  \left( \phi_H +
\frac{1}{\sqrt{N_2}} \xi_H\right )\Pi \right ]
\end{eqnarray}
The advantage of using the shift  operators relies in that they
admit a simple expansion in the limit for $N_1$ (resp. $N_2$) large:
\begin{eqnarray}
\label{ET}
 & \mathbb{E}^{\pm 1}_T = 1 \pm N_2^{-1/2} \frac{\partial}{\partial \xi_T} +
\frac{1}{2} N_2 ^{-1} \frac{\partial^2}{\partial \xi_T^2} \pm \cdots  \\
 \label{ERT}& \mathbb{E}^{\pm 1}_{R_T} = 1 \pm N_1^{-1/2}
\frac{\partial}{\partial
\xi_{R_T}} +\frac{1}{2} N_1^{-1} \frac{\partial^2}{\partial \xi_{R_T}^2} \pm
\cdots  \\
\label{EH}& \mathbb{E}^{\pm 1}_H = 1 \pm N_2^{-1/2}
\frac{\partial}{\partial\xi_H} +
\frac{1}{2} N_2 ^{-1} \frac{\partial^2}{\partial \xi_H^2} \pm \cdots
\end{eqnarray}
Plugging (\ref{ET})-(\ref{EH}) into (\ref{master_mod}), after  some
algebraic manipulation, one recovers at the leading order the
mean--field equations, formally identical to the ones reported
above, see equations (\ref{meanfield1})-(\ref{meanfield3}). The
next--to--leading order result in a Fokker Planck equation for the
probability distribution $\Pi(\underline{\xi}, t)$:
\begin{equation}
\label{FP}
 \frac{\partial \Pi}{\partial \tau}=-\sum_i \frac{\partial}{\partial \xi_i}
\left (A(\underline{\xi}) \Pi \right ) +\frac{1}{2} \sum_{ij}
B_{ij}\frac{\partial^2 \Pi}{\partial \xi_i \partial \xi_j}
\end{equation}
where:
\begin{equation*}
 A(\underline{\xi}) = \sum_j M_{ij}\xi_j
\end{equation*}
with
\begin{scriptsize}
\begin{displaymath}
\fl M = \left (
\begin{array}{ccc}
-\alpha(1-\phi_{R_T}^*) -\beta \phi_{R_T}^* -(\gamma+\sigma) \phi_H^* -\delta_1
-\eta_1  & c^{1/2}[\alpha \phi_T^*+\beta(1 -\phi_T^* -\phi_H^*)] & -\beta
\phi_{R_T}^*
-(\gamma+\sigma)\phi_T^* -\eta_1 \\
c^{1/2}[\alpha(1-\phi_{R_T}^*) +\beta\phi_{R_T}^*]  & -c[\alpha \phi_T^*+\beta(1
-\phi_T^* -\phi_H^*)] & c^{1/2}\beta \phi_{R_T}^*\\
\gamma \phi_H^* & 0 & \gamma \phi_T^* -\delta_2
 \end{array}
 \right)
\end{displaymath}
\end{scriptsize}
Notice that for $c=1$ (see \cite{MN2005}), $M$ reduces to the
Jacobian matrix (\ref{Jac}) which can be directly calculated from
the mean--field system. $B$ is instead a symmetric matrix whose
elements reads:
\begin{eqnarray*}
 \fl B_{11} & =  \alpha\phi_T^* (1-\phi_{R_T}^*) +\beta
\phi_{R_T}^* (1-\phi_T^*-\phi_H^*) +(\gamma+\sigma)
\phi_T^*\phi_H^*+\delta_1\phi_T^*+\eta_1(1-\phi_T^*-\phi_H^*)\\
 \fl B_{12} & =  -c^{1/2}[\alpha\phi_T^* (1-\phi_{R_T}^*) +\beta
\phi_{R_T}^* (1-\phi_T^*-\phi_H^*) ]\\
 \fl B_{13} & =  -\gamma  \phi_T^*\phi_H^*\\
 \fl B_{22} & =   c[\alpha\phi_T^* (1-\phi_{R_T}^*) +\beta
\phi_{R_T}^* (1-\phi_T^*-\phi_H^*) ]\\
 \fl B_{23} & =  0 \\
 \fl B_{33} & =  \gamma  \phi_T^*\phi_H^*+\delta_2 \phi_H^*
\end{eqnarray*}

\bibliographystyle{unsrt}
\bibliography{bibliography.bib}
\end{document}